\documentclass[journal=apchd5,manuscript=article]{achemso}
\usepackage[T1]{fontenc}
\usepackage{amsmath}
\usepackage{bbold}

\newcommand{\bv}[1]{\boldsymbol{#1}}
\newcommand{\tensor}[1]{\bar{\bar{#1}}}
\newcommand{\onlinecite}[1]{Ref.~[\hspace{-1 ex} \nocite{#1}\citenum{#1}]}

\author{Gilles Rosolen}
\email{gilles.rosolen@umons.ac.be}
\author{Bjorn Maes}
\affiliation[UMONS]
{Micro and Nanophotonic Materials Group, Research Institute for Materials Science and Engineering, University of Mons, Belgium}

\title{Strong multipolar transition enhancement with graphene nanoislands}

\keywords{High-order transition, Purcell enhancement, mode expansion, graphene, plasmonics}

\begin{document}

\begin{abstract}
During the past half century, a major approximation was natural in the field of light-matter interaction: the point-dipole model. It was assumed that the wavelength is much larger than the size of the emitting atom or molecule, so that the emitter can be described as a single or a collection of elementary dipoles. As it is legitimate for visible light, the approximation does no longer hold near plasmonic nanostructures, where the effective wavelength can drop below 10 nm. In that case deviations arise from the approximate model. First, the emitter spatial extent influences the far-field spectrum. Second, high-order transitions beyond the dipolar ones are not forbidden anymore. Going beyond the approximation requires intensive numerical efforts to compute the photonic response over the spatial extent of the emitter, since the complete Green's function is required. Here, we develop a general model that computes the multipolar transition rates of a quantum emitter in a photonic environment, by computing the Green's function through an eigenpermittivity modal expansion. We apply the method on graphene nanoislands, and we demonstrate a local breakdown of the selection rules, with quadrupolar transition rates becoming 100 times larger than dipolar ones.
\end{abstract}

\section{Introduction}

Usually the quantum emitter is approximated by a point emitting dipolar electric field. This is a legitimate approximation when the wavelength of the emitted light is much larger than the size of the atom or molecule. In that case, high-order transitions (transitions beyond the dipolar one, such as two-photon processes, electric and magnetic multipolar transitions...) are negligible and termed forbidden. However, those transitions are of particular importance in spectroscopy, photochemistry, quantum information and many other fields. Here, we develop a numerical method that computes the multipolar transition rates of a quantum emitter in a photonic environment, and we show that graphene nanoislands can break locally the conventional selection rules of a hydrogen-like emitter.

The transitions are governed by the coupling between the charged constituents of the emitter and the electromagnetic field. In free space, the latter is a plane wave, with expansion $\exp(i\bv{k}\cdot\bv{x}) \approx 1 + i(\bv{k}\cdot\bv{x}) - 0.5(\bv{k}\cdot\bv{x})^2 + ...$ in the limit of $\bv{k}\cdot\bv{x} \rightarrow 0$. Through the Fermi Golden rule, each term, which corresponds to a number of gradients of the electromagnetic field (0, 1, 2,...), can be traced back to a particular multipolar transition~\cite{Sanders2018}. For visible light, the wave vector $|\bv{k}|=10^7$ m$^{-1}$ and for hydrogen-like atoms $\langle |\bv{x}| \rangle\approx 10^{-10}$ m. It directly shows that the first term in the Taylor expansion, attributed to the dipolar transition, dominates by 3 orders of magnitude the linear order, attributed to the quadrupolar transition, and by 6 orders of magnitude the quadratic term, attributed to the octupolar transition~\cite{Filter2012}. Therefore, stronger field gradients over the spatial extent of the wavefunction of the emitter are necessary to enhance higher-order transitions~\cite{Sanders2018}.

One route toward non-negligible higher-order terms is therefore to consider larger emitters, as shown for quantum dots~\cite{Qian2019} and Rydberg excitons~\cite{Konzelmann2019}. Another route consists in enhancing the wave vector magnitude $|\bv{k}|$ by confining light in a nanophotonic structure. The wave vector can be written as $\bv{k}=\eta_0\omega/c$, with the confinement factor $\eta_0$ the ratio between the vacuum and effective wavelength. In this case, higher-order transitions are enhanced by a factor $\eta_0$ to the power of the considered order (for example the octupolar transition is enhanced by a factor $\eta_0^2$)~\cite{Rivera2016}. 

Under these conditions, plasmonic nanoantennas are ideal candidates to enhance higher order transitions. For instance, in noble metals, forbidden quadrupolar transitions are enhanced for emitters close to tips~\cite{ZuritaSnchez2002}, interfaces~\cite{Tojo2005}, nanowires~\cite{Rukhlenko2009}, nanogaps~\cite{Kim2018}, arrays~\cite{Yannopapas2015} and gold dimers~\cite{Filter2012,Neuman2018,Kosik2020}. The strongly confined graphene plasmons ($\eta_0 = 150-300$, depending on the absorption losses~\cite{Liu2008,Woessner2014}) form an excellent platform for high-order transitions, which can occur efficiently, even similar to dipolar transitions~\cite{Rivera2016}. In case of extremely high confinement ($\eta_0 > 500$) of plasmons in a two-dimensional material sheet, higher-order transitions rates can surpass lower order transitions, hence breaking the conventional selection rules~\cite{Rivera2016,Rivera2017}. 

Generally, accessing high-orders allows to probe a much larger range of the electronic energy level structure of an emitter, finding a way to a multiplex and broadband spectroscopy platform~\cite{Rivera2016,Neuman2018}. These higher-order transitions already play an important role in  spectroscopy of many relevant chemical species, from individual atoms~\cite{Tojo2004,EnzongaYoca2017} to larger molecules with high symmetry, such as dihydrogen, carbon dioxide, methane, and benzene~\cite{Li2011,Cheng2012}. In photochemistry, enhancing the magnetic dipole transition in oxygen is interesting for photochemical reactions~\cite{Manjavacas2017}. Finally, interference effects between multipolar orders can occur: the possibility of complete suppression of a certain transition through interference is required for many applications in the context of quantum computing, quantum storage, and quantum communication~\cite{Rusak2019,Qian2019}.

Despite its high potential and these developments, the field is currently limited by the difficulty of computing the electromagnetic environment of the emitter. Indeed, computing the spontaneous emission rates of a quantum emitter requires the knowledge of the electromagnetic field profile over the spatial extent of the wavefunctions of the emitter~\cite{Flick2018}. Usually, the problem is solved for absorption rates: in that case a plane wave excites a nanophotonic structure and the near-field is extracted~\cite{Kim2018, Sanders2018,ZuritaSnchez2002,Yannopapas2015}. This is a straightforward routine for conventional numerical methods such as the finite-element method~\cite{Lavrinenko2018,compphot2017} or the finite-difference time-domain method~\cite{allentaflove2013}. For spontaneous emission however, the knowledge of the vacuum field is essential. As a first approximation, one can resort to symmetric problems~\cite{Muniz2020}, or consider only the relevant (properly quantized) modes of the structure for the process~\cite{Filter2012,Rukhlenko2009,Neuman2018}. The complete resolution however requires knowledge of the Green's function, which is analytical only for uniform media and for simple geometries~\cite{Rivera2016,Rivera2017}. Numerical evaluation is very demanding with conventional numerical methods, as repeated simulations for different positions and orientations of a point dipole source are necessary~\cite{Kosik2020}.

In order to compute advanced photonic structures, a modal-based approach is very useful: a single simulation that determines the modes (e.g.\ of a cavity) is required to know the full spatial variation of the Green’s function~\cite{Lalanne2018}. The eigenpermittivity mode expansion is particularly suited for the spontaneous emission of an emitter, for which the emission frequency is fixed. Eigenpermittivity modes have a permittivity eigenvalue that pertains only to a scattering element, which spans a finite portion of space. As a result, the normalization is trivial. Furthermore they are orthogonal and appear to form a complete set~\cite{Bergman_Stroud_PRB_1980,Chen2019}. Once computed for a scatterer at a fixed wavelength, they straightforwardly give the optical response for any material constituting this scatterer. These modes have been derived in the seventies in the quasi-static approximation, and were used to derive bounds for scattering problems~\cite{Miller2016}, to study spasers~\cite{Bergman2003}, disordered media~\cite{Stockman:2004ey}, and second harmonic generation~\cite{Li2005}. The formalism (called GENOME for GEneralized NOrmal Mode Expansion) was recently extended beyond the quasistatic approximation by computing the electromagnetic fields and the associated Green’s function of open and lossy electromagnetic systems, in particular for general nanoparticle configurations using a commercial software (COMSOL Multiphysics)~\cite{Chen2019}. 
In this paper, we derive a general method to compute the transition rate of a quantum emitter influenced by its electromagnetic environment in the weak coupling regime (Section~\ref{sec:method}). The electromagnetic environment is embodied by the Green's function (computed with GENOME) and the quantum emitter is described by its wavefunction (a hydrogen-like emitter). Then, we apply this method to compute the electric dipolar, quadrupolar and octupolar transition rates of the emitter in the vicinity of graphene nanoislands for different geometries (triangle, square and crescent), showing strong enhancement of the transition rates (Section~\ref{sec:trans_rates}). Afterwards, we show that the graphene doping can be tuned to select particular transitions in Section~\ref{sec:tunability}, before demonstrating a local breakdown of the selection rules in Section~\ref{sec:breakdown}.

\section{Method} \label{sec:method}

We consider the spontaneous emission of atomic hydrogen-like emitters into plasmons given by the minimal coupling Hamiltonian~\cite{Scheel2008,craig1998molecular}
\begin{align}
    H &= H_\mathrm{a} + H_\mathrm{em} + H_\mathrm{int} \\
    H_\mathrm{a} &= \sum_i \frac{\bv{p}_i^2}{2m_e} - \frac{e^2}{4\pi\varepsilon_0 r} + H_\mathrm{e-e} + H_\mathrm{SO} \\
    H_\mathrm{em} &= \sum_{j=x,y,z} \int d \bv{r} \int d\omega \hbar \omega \left[ f_j^\dagger(\bv{r},\omega) f_j(\bv{r},\omega) + \frac{1}{2}\right] \\
    H_\mathrm{int} &= \sum_i \frac{e}{2m_e} \left( \bv{p}_i \cdot \bv{A}(\bv{r}_i) + \bv{A}(\bv{r}_i) \cdot  \bv{p}_i \right) + \frac{e^2}{2m_e} \bv{A}^2(\bv{r}_i) + \frac{e\hbar}{2m_e}\sigma_i \cdot \bv{B}(\bv{r}_i)
\end{align}
with $\bv{p}_i$, $\bv{r}_i$ and $\sigma_i$ the impulsion, position and spin of the $i^\mathrm{th}$ electron, $e$ the electronic charge, $m_e$ the electron mass, $\bv{A}$ and $\bv{B}$ the vector potential and magnetic field, $H_\mathrm{SO}$ the spin-orbit coupling, and $H_\mathrm{e-e}$ the electron-electron interaction. $f_j^\dagger(\bv{r},\omega)$ and $f_j(\bv{r},\omega)$ are the creation and annihilation operators, respectively.

For the interaction Hamiltonian, we neglect the ponderomotive potential ($\bv{A}^2$ term) and the $\bv{B}$ term as the latter is negligible for non-magnetic structures~\cite{Sloan2019}. We also neglect the $\bv{p}\cdot\bv{A}$ term since we use the Coulomb gauge (the scalar potential is identically 0) and therefore $\nabla \cdot \bv{A} = 0$ except at an interface: With the atom-interface distance we consider, and the rapid decay of the atom wavefunctions, the contribution of this term will be negligible. Writing the vector potential with the Green's function of the system, and applying the Fermi Golden rule, one finds (for details, see~\cite{Rivera2016})
\begin{align}
    \Gamma = \frac{2\pi}{\hbar^2} \frac{e^2\hbar^3}{\pi\varepsilon_0 m_e^2 c^2} \int \int d\bv{r} d\bv{r}' \psi_e^*(\bv{r}) \psi_e(\bv{r}') \bv{\nabla} \psi_g(\bv{r}) \cdot \mathrm{Im} \tensor{G}(\bv{r},\bv{r}',\omega_0) \cdot \bv{\nabla} \psi_g^*(\bv{r}') \label{Eq:Gen_rate}
\end{align}
where $\varepsilon_0$ is the vacuum permittivity, $c$ is the speed of light, $\psi_g$ and $\psi_e$ are the atomic wavefunctions of the ground and excited state of the emitter. $\tensor{G}(\bv{r},\bv{r}',\omega_0)$ is the Green's function of the Maxwell equations and satisfies $\nabla \times (\nabla \times \tensor{G}) - \frac{\omega^2}{c^2}\varepsilon_r(\bv{r},\omega)\tensor{G} = \tensor{I} \delta(\bv{r}-\bv{r}')$, with $\varepsilon_r$ the relative permittivity and $\delta$ the delta function~\cite{Novotny2006}.

Equation~\ref{Eq:Gen_rate} computes the transition rates of any emitter (via the atomic wavefunction) within any photonic environment (described by the Green's function) in the weak coupling regime. As mentioned, the spatial variation of the Green’s function is known analytically for uniform media and for simple geometries. However, more complex structures need to be evaluated numerically, with high computational cost~\cite{Rosolen2020}. Equation~\ref{Eq:Gen_rate} is also resource demanding since the integration is performed over 6 dimensions ($\bv{r}$ and $\bv{r}'$). In order to resolve these two issues, we resort to GENOME~\cite{Chen2019}, with the advantage that one modal computation allows the knowledge of the complete spatial Green's function.

In GENOME, the problem is written at a fixed frequency, and the mode-related eigenvalue is the permittivity. This formulation suits well the determination of spontaneous emission rates since the emission frequency is determined by the emitter. The modes $\bv{E}_m(\bv{r})$ of the scatterer are solved with a commercial finite-element based software (COMSOL Multiphysics)~\cite{Rosolen2020} and the Green's function becomes
\begin{align}
\tensor{G}(\bv{r}, \bv{r}') = \tensor{G}_0(\left|\bv{r}- \bv{r}'\right|) + \frac{1}{k^2} \sum_m \frac{\varepsilon_i - \varepsilon_b}{(\varepsilon_m - \varepsilon_i)(\varepsilon_m - \varepsilon_b)} \bv{E}_m(\bv{r}) \otimes \bv{E}_m^\dagger(\bv{r}'), \label{Eq:Green}
\end{align}
where $m$ is the mode number, $k$ is the vacuum wave vector, $\varepsilon_m$ is the eigenpermittivity, $\varepsilon_i$ is the permittivity of the scatterer, $\varepsilon_b$ the permittivity of the background material and $\tensor{G}_0(\left|\bv{r}- \bv{r}'\right|)$ the Green's function of vacuum, which has an analytical form~\cite{Novotny2006}.

Inserting Equation~\ref{Eq:Green} in Equation~\ref{Eq:Gen_rate}, we immediately see that the rate is a sum of two contributions $\Gamma = \Gamma_0 + \Gamma_s$, with $\Gamma_0$ the decay rate in vacuum (based on the contribution of $\tensor{G}_0(\left|\bv{r}- \bv{r}'\right|)$), and $\Gamma_s$ depending on the modes and hence, the nanophotonic structure. Focusing on this $\Gamma_s$ contribution, we can write
\begin{align}
    \Gamma_s = \frac{2\pi}{k^2} \frac{e^2 \hbar}{\pi \varepsilon_0 m_e^2 c^2} \int \int d\bv{r}d\bv{r}' \psi_e^*(\bv{r})\psi_e(\bv{r}') \bv{\nabla_r} \psi_g(\bv{r}) \left( \mathrm{Im} \sum_m \gamma_m \bv{E}_m(\bv{r}) \otimes \bv{E}_m^\dagger(\bv{r}') \right) \bv{\nabla}_{\bv{r}'}\psi_g^*(\bv{r}') 
\end{align}
where we defined $\gamma_m=\frac{\varepsilon_i - \varepsilon_b}{(\varepsilon_m - \varepsilon_i)(\varepsilon_m - \varepsilon_b)}$. Note that the adjoint field ($\bv{E}^\dagger$) is the transposed vector, and there is no complex conjugate~\cite{Chen2019}.

Since we can choose the wavefunctions to be real, the complex conjugate for the wavefunctions disappears, and we can integrate separately for $\bv{r}$ and $\bv{r}'$. Both integrations give the same value, leading to:
\begin{equation}
    \Gamma_s = \frac{2\pi}{k^2} \frac{e^2 \hbar}{\pi \varepsilon_0 m_e^2 c^2} \sum_m  \mathrm{Im} \left[  \gamma_m \left( \int \psi_e(\bv{r}) \bv{E}_{m}(\bv{r}) \bv{\nabla_r} \psi_g(\bv{r})  d\bv{r} \right)^2 \right] \label{Eq:rate_final}
\end{equation}
Finally, the transition rates are obtained with a 3 dimensional integration over the wavefunctions and the modes profiles, with a sum that can be truncated once the convergence is sufficient (40 modes in our case, see Supplementary Information). Note that when the integral is computed, the rate can be known for any material constituting the scatterer, enclosed in the parameter $\gamma_m$. In that regards, graphene is the perfect candidate as it can be tuned to match a particular resonance (see Section~\ref{sec:tunability}).

In this work the graphene nanoislands are modeled with an effective thickness of $t=1$~nm. The graphene permittivity ($\varepsilon_i$) is deduced from the surface optical conductivity ($\sigma=\widetilde{\sigma}_{\mathrm{intra}} + \widetilde{\sigma}_{\mathrm{inter}}$) with $\varepsilon_i = 1+i\sigma/\omega\varepsilon_0 t$. The optical conductivity is derived within the local random-phase approximation model \cite{Falkovsky2007,Falkovsky2008} and is the sum of the two following contribution:
	\begin{align}
			\widetilde{\sigma}_{\mathrm{intra}}&=\frac{2 i e^2 k_B T}{\hbar^2 \pi (\omega+i \tau_g^{-1})} \ln \left[2 \cosh \left( \frac{E_F}{2k_B T} \right) \right] \label{sigma_intra} \\ 
			\widetilde{\sigma}_{\mathrm{inter}}&=\frac{e^2}{4 \hbar} \left[ \frac{1}{2} + \frac{1}{\pi} \arctan \left( \frac{\hbar \omega - 2 E_F}{2k_B T} \right) \right]- \frac{e^2}{4 \hbar} \left[\frac{i}{2 \pi} \ln \frac{(\hbar \omega + 2 E_F)^2}{(\hbar \omega - 2 E_F)^2 + (2 k_B T)^2} \right]  \label{sigma_inter} 
	\end{align}
with $T$ the temperature, $k_B$ the Boltzmann constant and $E_F$ the doping level of graphene. The scattering lifetime of electrons in graphene depends on the doping and is given by $\tau_g=\mu E_F/e v_F^2 \approx 10^{-12}$~s for $E_F=1$~eV, with the impurity-limited DC conductivity $\mu\approx 10000$~cm$^2$/(V s) and $v_F=10^6$~m/s the graphene Fermi velocity~\cite{Christensen2012,Novoselov2004a}.

The integration of Equation~\ref{Eq:rate_final} is successfully compared to direct simulations of dipolar and quadrupolar transitions in the SI, showing great convergence with only 40 modes (1\% relative error). In the next section, we implement Equation~\ref{Eq:rate_final} to compute the rate of a H-like atom close to graphene nanoislands of varying geometry.

\section{Results and discussion} 

We apply our method to compute the electric dipolar (E1), quadrupolar (E2) and octupolar (E3) transition rates of a H-like atom close to a graphene sheet with triangle, square and crescent geometry. We consider the transition series 6{p, d, f} $\rightarrow$ 4s, which are E1, E2 and E3 transitions, respectively. Note that we suppose that the angular magnetic number remains $m=0$ during the transition, and we rotate the emitter wavefunctions to match the corresponding classical point-dipole orientation. The free-space wavelengths of the transitions are all $\lambda = 2.63$~$\mu$m, and in the whole paper the emitter is situated 5~nm above the graphene surface.

We then discuss the rate dependence on the graphene doping (Section~\ref{sec:trans_rates}) and we demonstrate the advantage of graphene tunability for multipolar transitions (Section~\ref{sec:tunability}). Finally, we optimize a configuration where the conventional selection rules break down, i.e.\ when the quadrupolar transition rate dominates the dipolar one (Section~\ref{sec:breakdown}).

\subsection{Transition rates} \label{sec:trans_rates}

\begin{figure*}[t!]
     \centering
     \includegraphics[width=1.0\textwidth]{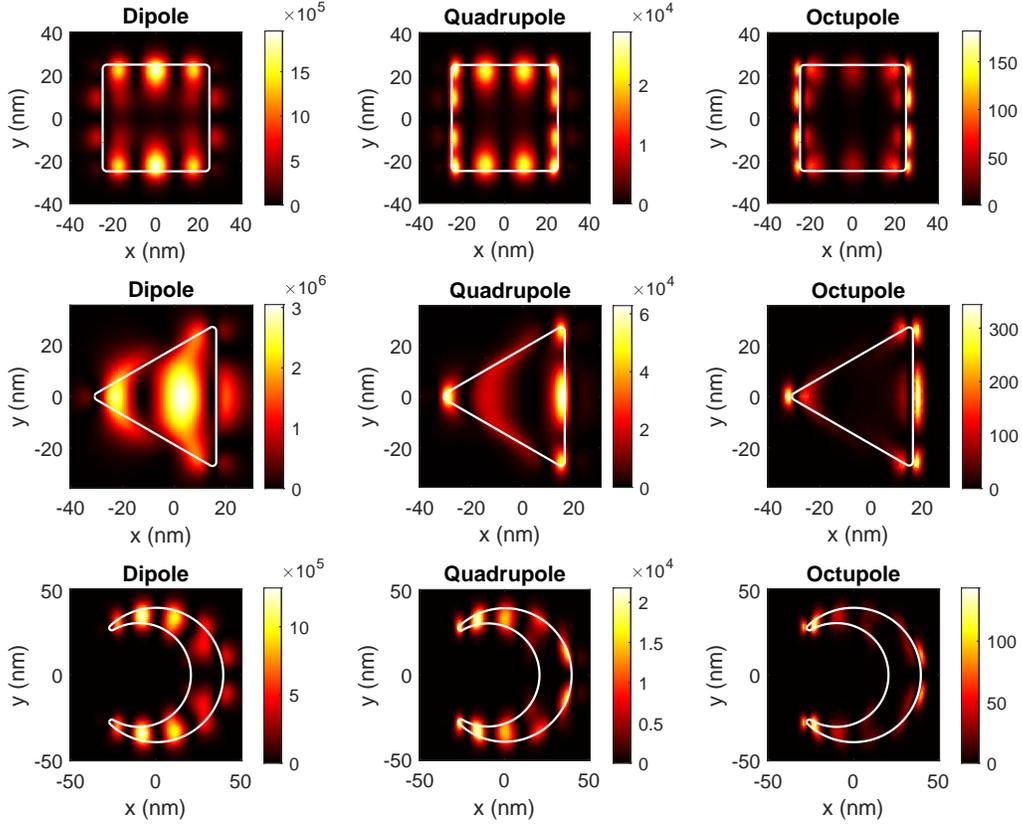}
     \caption{\textbf{Dipolar, quadrupolar and octupolar transition rates of an $x$-oriented emitter close to graphene nanoislands of various geometries.} The dipolar (left), quadrupolar (center) and octupolar (right) transition rate as a function of the emitter position, which is 5~nm above the graphene nanoislands: 50~nm side length square (up), 50~nm side length triangle (middle), and 80~nm height crescent (bottom). The geometry boundaries are represented with a solid white line, and the rates are normalized by the dipolar emission rate in free space $\Gamma_0$. For the triangle $E_F=0.98$~eV, for the square $E_F=0.72$~eV, and for the crescent $E_F=0.88$~eV.} \label{fig:trans_rates}
 \end{figure*}

Figure~\ref{fig:trans_rates} shows the dipolar ($\Gamma_D$), quadrupolar ($\Gamma_Q$) and octupolar  ($\Gamma_O$) transition rates of a H-like emitter in the vicinity of a graphene nanoisland for three geometries: square, triangle and crescent shape. The rates are normalized by the dipole emission rate in free space $\Gamma_{D,0} = 4.484\times 10^5$~s$^{-1}$. The latter is obtained by integrating Equation~\ref{Eq:Gen_rate} in free space, and is in perfect agreement with the experimental values~\cite{Wiese2009} (more information in SI). One can see, for example, that the octupolar rate is strongly enhanced with respect to vacuum: it is up to 300 times stronger than the dipolar rate in free space for the triangle geometry.

One observes that the strongest quadrupolar and octupolar rate enhancement appears along the edges and corners of the geometries. This is a consequence of the strong field gradients appearing along the graphene edge~\cite{Sanders2018,Karanikolas2019}. Second, for all geometries, the maximum quadrupolar rate is two orders of magnitude smaller than the maximum dipolar rate. That two-order magnitude difference compares fairly with the rates comparison obtained in \onlinecite{Rivera2016} for a H-like emitter close to a non-structured two-dimensional material supporting plasmons confined with a factor $\eta_0 \approx 35 - 50$ (corresponding to doping between 0.7 and 1~eV). The four-order magnitude difference between the dipolar and octupolar transition rates is also in agreement with the literature.

With the graphene nanoislands, we break the in-plane translational symmetry and the conventional dominance of the dipolar transition rate over the quadrupolar transition rate. From the spatial maps, we observe that the maxima of the quadrupolar rate do not coincide with the maxima of the dipolar rate: by moving the emitter, one can find a position where the quadrupolar rate dominates the dipolar rate, breaking the conventional selection rules (see Section~\ref{sec:breakdown}). 

Note that the $z$-oriented emitter (out-of-plane direction) shows stronger rate enhancement, but the dipolar, quadrupolar and octupolar transitions are all three enhanced at the same position of the $z$-oriented emitter (see SI). This originates from the field profile of the modes at the edge, showing a strong confinement in the $x-y$ plane, while having the confinement of a graphene sheet in the $z$-direction ($\eta_0 \approx 50$)~\cite{Nikitin2011,Sanders2018}. The latter implies the conservation of the selection rules for the $z$-oriented emitter for this confinement i.e.\ at these graphene doping levels~\cite{Rivera2016}.

\subsection{Graphene tunability} \label{sec:tunability}

In a spontaneous emission process the emission wavelength is fixed via the considered transition. Hence, as the frequency of the source is not a variable, a tuning knob is offered by the environment, e.g.\ the permittivity of the scatterer. The considered mode expansion is particularly well suited for this context as the permittivity is the eigenvalue of the problem. As a consequence, the permittivity of the scatterer only appears as a multiplicative constant of the three-dimensional integral ($\gamma_m$ in Equation~\ref{Eq:rate_final}), allowing to select the mode resonating with the transition.

\begin{figure*}[t!]
     \centering
     \includegraphics[width=1.0\textwidth]{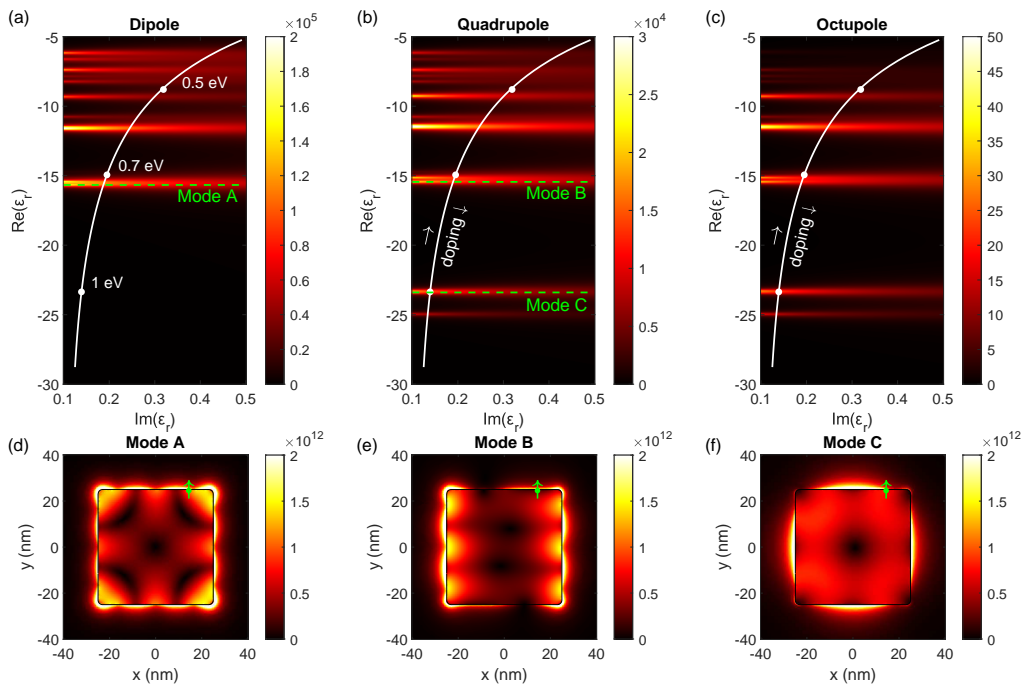}
     \caption{\textbf{Choosing the material of the scatterer to enhance particular transition rates.} The (a) dipolar, (b) quadrupolar and (c) octupolar transition rate enhancement in color scale as a function of the relative permittivity of the scatterer (real and imaginary parts). The rates are normalized by the dipolar emission rate in free space. The white line represents the permittivity range covered by graphene upon doping (white dots at 0.5 eV, 0.7 eV and 1 eV doping). The horizontal green dashed lines indicate the modes contributing the most to the considered transition. Their mode profiles ($|E|$) are represented in (d) $\varepsilon_m = -15.65$, (e) $\varepsilon_m = -15.45$ and (f) $\varepsilon_m = -23.4$. The green dot and arrow represent the position and orientation of the emitter.} \label{fig:permGraph}
 \end{figure*}

Figures~\ref{fig:permGraph}a, b and c show the transition rates' dependence on the permittivity of the material, for a square two-dimensional material with variable permittivity and with side length of 50 nm. The emitter is $y$-oriented at the position (14.4;24.5) nm, 5 nm above the material (green arrow in Figures~\ref{fig:permGraph}d, e and f). The transition rate map is characterized by horizontal lines of enhanced transition rates. Each line directly corresponds to a plasmonic mode of the structure, which constitutes the dominant decay route for this transition. For example, the dipolar transition couples with the mode A (represented in Figure~\ref{fig:permGraph}d) and the quadrupolar and octupolar transitions couple with modes B and C (represented in Figure~\ref{fig:permGraph}e and f). Note that modes A and B are different, but their eigenvalues are close (resp.\ $\varepsilon_m = -15.65$ and $\varepsilon_m = -15.45$). Their proximity implies that the dipolar rate ($\Gamma_D/\Gamma_{D0} = 1.1\times 10^5$) dominates the quadrupolar rate ($\Gamma_Q/\Gamma_{D0} = 1.5\times 10^4$). 

Mode C is more interesting since it couples strongly with the quadrupolar transition ($\Gamma_Q/\Gamma_{D0} = 1.2\times 10^4$) and weakly with the dipolar transition ($\Gamma_D/\Gamma_{D0} = 0.6\times 10^3$), comparing Figures~\ref{fig:permGraph}a and b. This conclusion corresponds with the field profile in Figure~\ref{fig:permGraph}f at the position of the dipole (green dot). The emitter is placed at a position where the field has a low value (weak enhancement of dipolar transition), but near the edge, where the field gradient is the strongest (strong enhancement of quadrupolar transition).) 

In the horizontal direction of Figures~\ref{fig:permGraph}a, b and c the imaginary part of the permittivity of the material broadens the resonance peak, consequently reducing the maximum value of the transition rate. 

Therefore, selecting a square two-dimensional material (of this size, and at $\lambda = 2.63$~$\mu$m) for a particular permittivity can enhance a particular transition. For example, in order to produce an electric octupolar transition rate 50 times stronger than the transition dipolar rate in free space, one can choose a material of real relative permittivity of -23.4.

The white lines in Figures~\ref{fig:permGraph}a, b and c represent the permittivity range covered by graphene at this wavelength. A fine tuning of the graphene doping thus allows to select the plasmonic mode that will dominate the transition, and hence the transition order. In the following section, we show a particular doping of graphene where the quadrupolar rate dominates the dipolar transition rate, consequently breaking the conventional selection rules.

\subsection{Local breakdown of conventional selection rules} \label{sec:breakdown}

At particular positions of the emitter, the quadrupolar transition rate overcomes the dipolar transition rate. This breakdown occurs at ultra-strong plasmon confinement ($\eta_0 > 500$) for planar two-dimensional materials~\cite{Rivera2016}, which is experimentally achievable with graphene, but at the cost of considerable absorption losses~\cite{Liu2008}. The shape of the graphene nanoislands provides another degree of freedom to mold the field profile and break the selection rules.

\begin{figure*}[t!]
     \centering
     \includegraphics[width=0.8\textwidth]{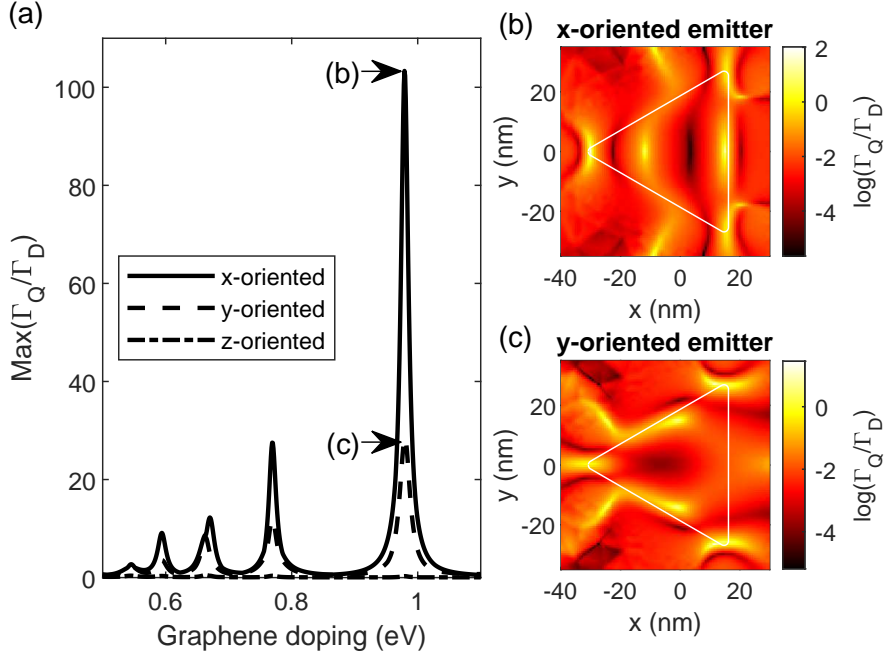}
     \caption{\textbf{Local breakdown of the selection rules with the triangle graphene nanoisland.} (a) The maximum of the quadrupolar rate ($\Gamma_Q$) over the dipolar rate ($\Gamma_D$) for a H-like emitter 5 nm above the triangular graphene nanoisland, evaluated for varying graphene doping and emitter orientation. The $z$-oriented emitter does not demonstrate a breakdown of the selection rules (ratio always smaller than 1). A maximum is obtained for a graphene doping of $E_F = 0.98$ eV. At this doping, (b) shows the logarithmic value of the ratio for an $x$-oriented emitter, and (c) a $y$-oriented emitter. The boundaries of the triangular graphene nanoisland are represented with a solid white line.} \label{fig:QuadoverDip}
 \end{figure*}
 
We focus on the triangular graphene nanoisland of 50 nm side length, for which we computed the normalized dipolar and quadrupolar rates of an emitter 5 nm above its surface (Figure~\ref{fig:trans_rates}). In Figure~\ref{fig:QuadoverDip}a, we plot the maximum of the ratio $\Gamma_Q/\Gamma_D$ over all positions of the emitter, as a function of the graphene doping. This shows that the quadrupolar transition rate can be up to 100 times stronger than the dipolar transition rate at particular positions, breaking locally the conventional selection rules (the value is converged for 40 modes, as shown in the SI). 

In Figures~\ref{fig:QuadoverDip}b and c, for the $x$- and $y$-oriented emitter, respectively, we observe enhancement where the field demonstrates strong gradients, i.e.\ at the corner of the triangle or along the edge. On the contrary, as observed in Figure~\ref{fig:QuadoverDip}a for the $z$-oriented emitter, the dipolar rate always dominates the quadrupolar one (the maximum rate enhancement of each order appear at the same position, as discussed in Section~\ref{sec:trans_rates}). 

Note that the maximum is not a consequence of an inhibited dipolar transition: the quadrupolar rate is strongly enhanced. For example, for an $x$-oriented emitter (Figures~\ref{fig:QuadoverDip}b) at the left corner of the triangle, the dipolar transition remains enhanced ($\Gamma_D/\Gamma_{D,0}=0.51\times 10^3$), but its rate is weaker than the quadrupolar rate, which is $5.3\times 10^4$ times the dipolar transition in free space.

Other areas further away from graphene seem to demonstrate a strong quadrupolar enhancement (for example position (-9;35) nm in Figure~\ref{fig:QuadoverDip}b). However, these are regions where the dipolar transition is poorly enhanced ($\Gamma_D/\Gamma_{D,0} = 16$), as well as the quadrupolar transition ($\Gamma_Q/\Gamma_{D,0} = 161$).

\section{Conclusions and perspectives}

We develop a numerical method that allows the computation of the transition rates of an emitter in the vicinity of any scatterer in the weak coupling regime, with the prerequisite knowledge of the wavefunctions of the excited and ground states of the emitter. With this method two important parameters are easily variable: the position of the emitter and the permittivity of the material, the latter being easily tunable for graphene. By optimizing both parameters we demonstrate a breakdown of the selection rules, with the quadrupolar transition rate 100 times stronger than the dipolar transition rate for a H-like emitter in the vicinity of a triangular graphene nanoisland. These results uncover interesting perspectives for applications in spectroscopy, photochemistry and quantum technologies.

In this paper, we focus on a particular photonic structure, i.e.\ a scatterer in free space. For future exploration, GENOME also allows the determination of the Green's function for more complex structures, such as a scatterer on a substrate~\cite{Chen2019}, non-uniform scatterers~\cite{Chen2020}, cluster of scatterers and finite periodic structures~\cite{Rosolen2020}. For experimental observation the investigation of these designs is important for the coupling of produced photons to the far-field. As an example, combining the near-field results (e.g.\ Figure~\ref{fig:permGraph}) with the far-field out-coupling efficiency of the dominant mode, allows to select the graphene doping necessary to reach sufficient far-field emission. Such an analysis was carried out e.g.\ for two-photon emission processes near graphene nanoislands~\cite{Muniz2020}. Other structures may be envisaged to enhance the coupling of a plane wave with a quadrupolar transition~\cite{Sakai2018}.

In parallel, our method allows for the computation of larger atoms and complex molecules by combining GENOME with time-dependent density functional theory techniques. Hence, controlling the emission rate of quantum dots~\cite{ZuritaSnchez2002,Qian2019} and Rydberg excitons~\cite{Konzelmann2019} in complex electromagnetic environments is within reach. Furthermore, since the dipolar and quadrupolar rates compete, destructive interference effects can be observed and lead to suppression of particular transition channels~\cite{Rusak2019,Qian2019}, leading to diverse quantum applications such as quantum computing, quantum storage, and quantum communication~\cite{Rusak2019,Qian2019}.

\bibliography{bibli}

\pagebreak
\begin{center}
\textbf{\huge Supplementary Information}
\end{center}
\setcounter{equation}{0}
\setcounter{figure}{0}
\setcounter{table}{0}
\setcounter{page}{1}
\setcounter{section}{0}
\makeatletter
\renewcommand{\theequation}{S\arabic{equation}}
\renewcommand{\thefigure}{S\arabic{figure}}
\renewcommand{\thesection}{S\arabic{section}}

In this supplementary file, we discuss the method implementation. First, the results' convergence is shown for the different parameters in the integration. A verification of the dipolar transition rate in free-space is also proposed. Second, the integration is compared to full wave simulations and shows good agreement. Third, results are given for the quadrupolar transition rate dominance over the dipolar one for the square graphene nanoisland. Finally, supplementary figures complete Figure~\ref{fig:trans_rates} of the main text.

\section{Convergence of the results}

In this section, we show the convergence of the integration, with a cut-off set for 1\% relative error. We implemented a spherical integration over the radial component and the azimuthal and polar angles. The wavefunctions are those of the hydrogen atom, and the field is obtained through an adjusted eigenfrequency study in COMSOL Multiphysics (see Ref.~\cite{Rosolen2020} of the main text). The extracted field is linearly interpolated through the Matlab Scattered Interpolant feature, requiring a fine mesh of 0.8 nm over the integration domain of the emitter for the octupolar transition.

\begin{figure*}[t]
     \centering
     \includegraphics[width=1\textwidth]{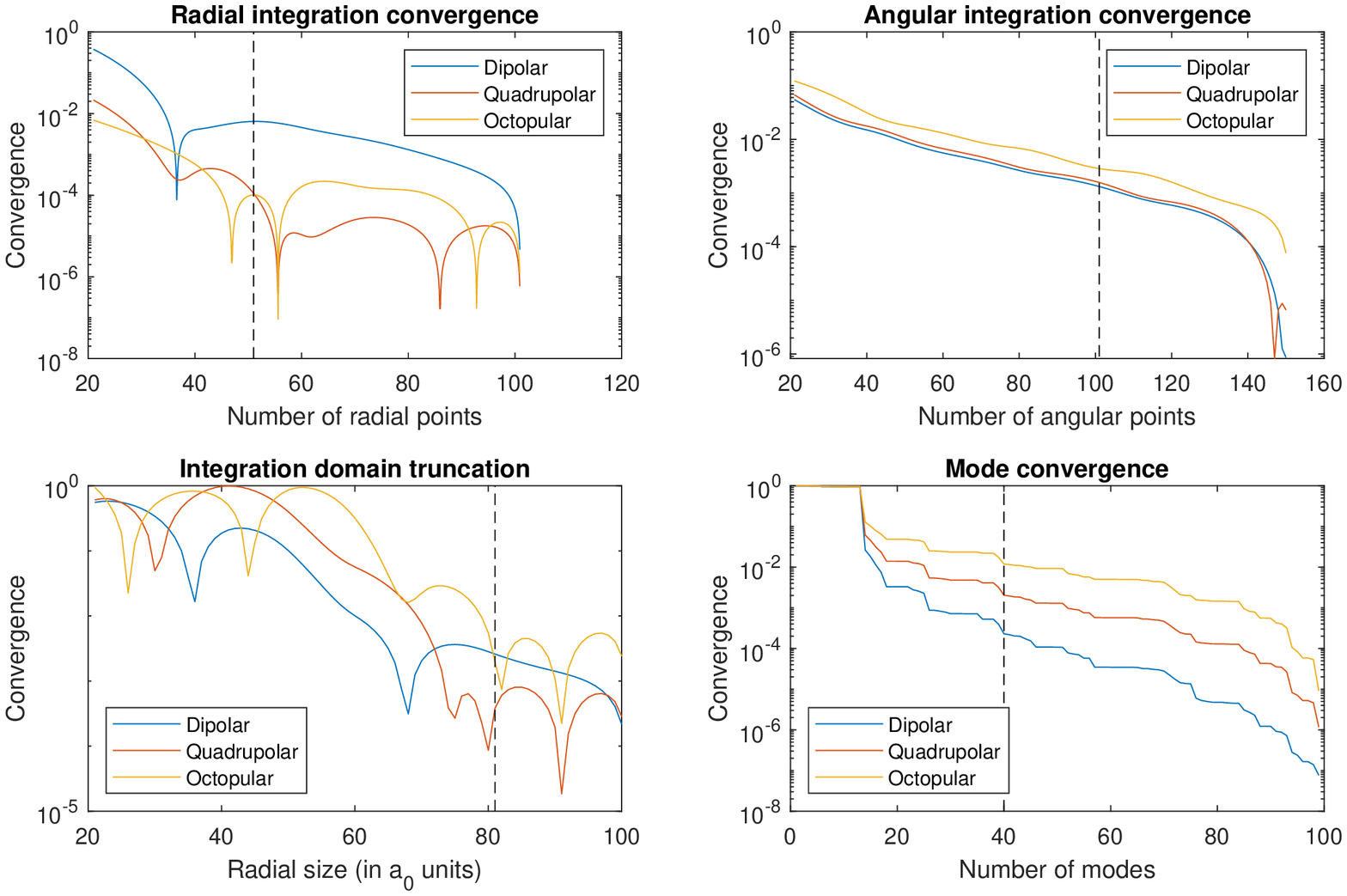}
     \caption{\textbf{Convergence of the integration for the radial and angular accuracy, the truncation of the integration domain, and the number of modes considered.} The convergence is checked for the transitions 6{p,f,d} $\rightarrow$ 4s of a $z$-oriented hydrogen atom 5 nm above the center of one edge of the square graphene nanoisland. The vertical dashed lines show the cut-off value used for the computations to reach 1\% accuracy.} \label{fig:SIconvergence}
 \end{figure*}

The convergence is checked as a function of the number of subdivisions in the angular direction and in the radial direction, see Figure~\ref{fig:SIconvergence}. A decent convergence is reached with 100 angular subdivisions and 50 radial subdivisions. The integration domain was truncated to $81a_0$ (with $a_0$ the Bohr radius), as it reaches convergence for the dipolar transition. Finally, 40 plasmonic modes are necessary to reach a 1\% relative error for the transition rate. Note that in Figure~\ref{fig:SIconvergence}, the convergence is reached for a graphene doping of 0.7~eV: when the graphene doping decreases below 0.5~eV, more modes are necessary to reach 1\% convergence (60 modes). 
 
\begin{figure*}[t]
     \centering
     \includegraphics[width=1\textwidth]{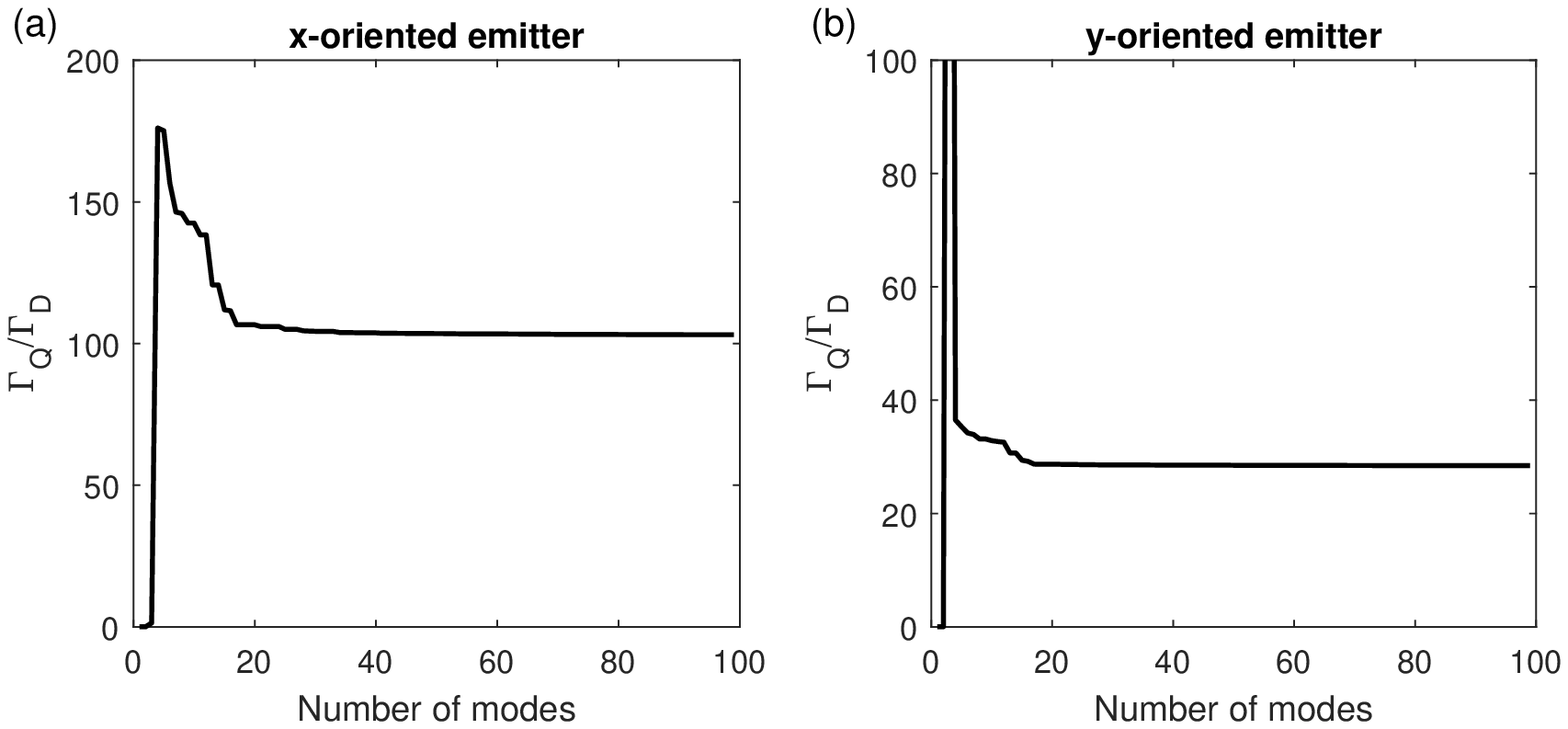}
     \caption{\textbf{Convergence of the integration showing that quadrupolar rates overcome dipolar rates.} $\Gamma_Q/\Gamma_D$ as a function of the number of modes for a triangle graphene nanoisland of $E_F=0.979$ eV for (a) an $x$-oriented emitter at the position (-30.2;0) nm and for (b) a $y$-oriented emitter at the position (11.1;26.6) nm.} \label{fig:SIconvQuandoverDip}
 \end{figure*}

Figure~\ref{fig:SIconvQuandoverDip} shows that the ratio $\Gamma_Q/\Gamma_D$ is also converged with 40 modes (so Figure 3 of the main text is a converged result). These two figures are obtained with the triangular graphene nanoisland, of 50-nm side length with the rotation center at the origin of our reference system. The two emitter orientations described in the main text ($x$ and $y$) are considered, each at a particular position. 

In this paper, all rates are normalized by the dipole emission rate in free space. Considering the long wavelength approximation in vacuum ($\lambda \gg a_0$, with $a_0$ the Bohr radius), and using that the imaginary part of the Green's tensor in vacuum at the position of the dipole is $\mathrm{Im} \left[ \tensor{G}_0(\bv{r},\bv{r},\omega_0) \right]= \mathbb{1} k/6\pi$, we have
\begin{equation}
\begin{aligned}
    \Gamma_0 &= \frac{2\pi}{\hbar^2} \frac{e^2\hbar^3}{\pi\varepsilon_0 m_e^2 c^2} \int \int d\bv{r} d\bv{r}' \psi_e^*(\bv{r}) \bv{\nabla}_{\bv{r}} \psi_g(\bv{r}) \cdot \mathrm{Im} \left[ \tensor{G}_0(\bv{r},\bv{r}',\omega_0)\right] \cdot \bv{\nabla}_{\bv{r}'} \psi_g^*(\bv{r}')\psi_e(\bv{r}') \\
    &= \frac{k}{3} \frac{e^2 \hbar}{\pi \varepsilon_0 m_e^2 c^2} \int \int d\bv{r}d\bv{r}' \psi_e^*(\bv{r})\psi_e(\bv{r}') \bv{\nabla_r} \psi_g(\bv{r})\cdot \bv{\nabla}_{\bv{r}'}\psi_g^*(\bv{r}') \\
    &=\frac{k}{3} \frac{e^2 \hbar}{\pi \varepsilon_0 m_e^2 c^2} \left| \int d\bv{r}\psi_e^*(\bv{r}) \bv{\nabla_r} \psi_g(\bv{r}) \right|^2  
\end{aligned}
\end{equation}
The integration leads to $\Gamma_0 = 4.484\times 10^5$ s$^{-1}$ in good agreement with the tabulated experimental values for the transition 6p $\rightarrow$ 4s ($\Gamma_D = 4.456\times 10^5$ s$^{-1}$), considering the 1\% error on the integration.

\section{Comparison with direct COMSOL simulations}

\begin{figure*}[t]
     \centering
     \includegraphics[width=1\textwidth]{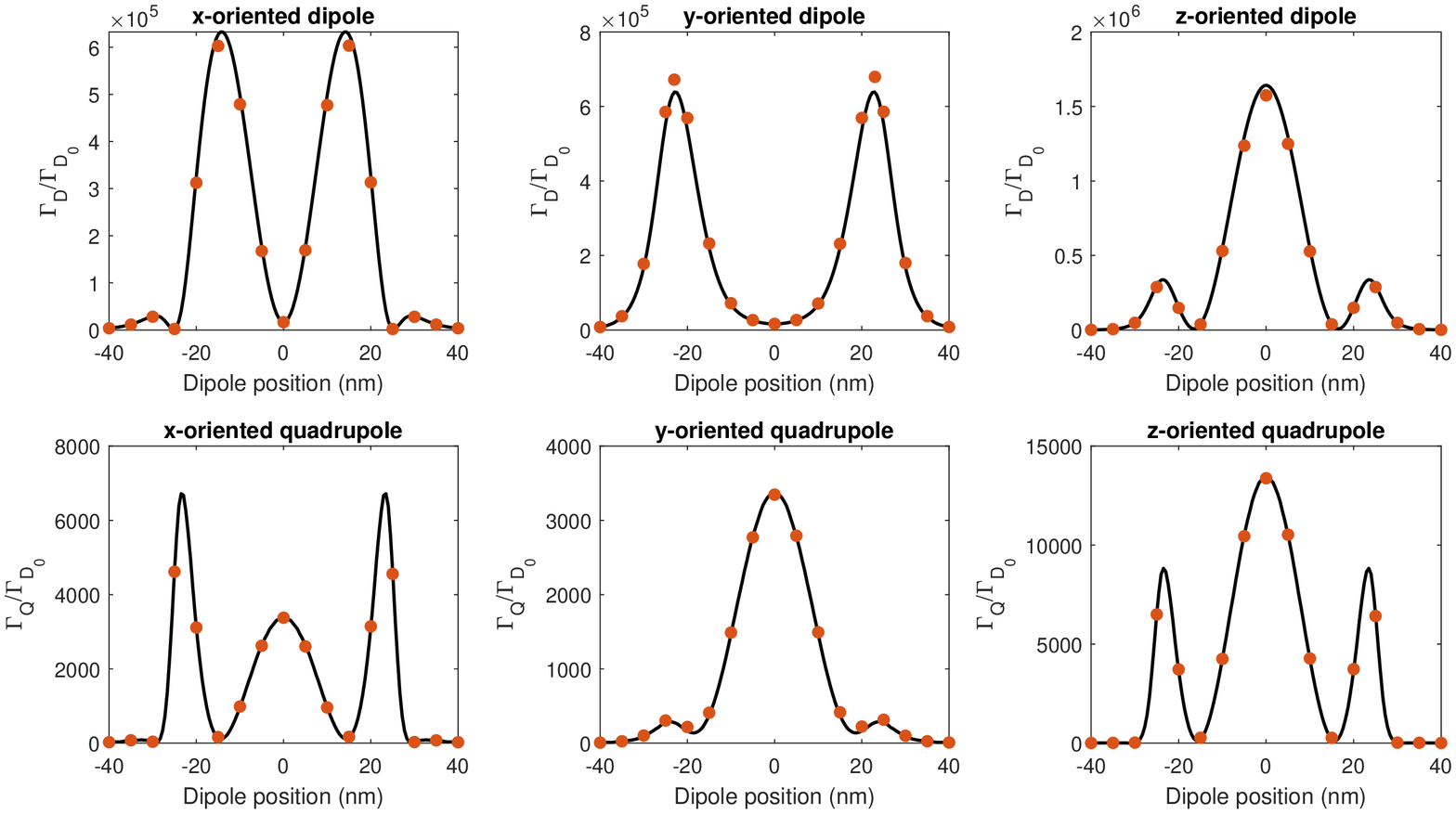}
     \caption{\textbf{Good agreement for the dipolar and quadrupolar rates obtained via direct COMSOL simulations and via mode integration based on the wavefunctions of the emitter.} The comparison is done for the square graphene nanoisland of 0.7 eV doping and with side length 50 nm. The emitter position follows the horizontal median of the square, 5 nm above the graphene surface. The points represent COMSOL direct simulations, while the solid line represents the integration.} \label{fig:SIcomp_com_dip_quad}
 \end{figure*}

The comparison is conducted for the square graphene nanoisland. We compare the dipole and quadrupole transition rates evaluated with COMSOL direct simulations and with Equation~\ref{Eq:rate_final} of the main text. In COMSOL, the structure is a square of 50 nm side length and 1 nm thickness, with rounded corners and maximum element mesh size of 1 nm. The dipole is represented by a point and is enclosed in a sphere of radius 2 nm that integrates the Poynting vectors normal to the surface (in this small region the mesh elements have a typical size of 0.3 nm). The obtained value is normalized by the value calculated through the exact same simulation in vacuum, providing the dipole emission rate in the top of Figure~\ref{fig:SIcomp_com_dip_quad}. The comparison between COMSOL (orange dots) and with Equation~\ref{Eq:rate_final} of the main text (solid line) is good.

The quadrupolar rates are computed with two dipoles separated by 0.3 nm in opposition and in phase. The rates are then normalized by the dipolar rate in free space, providing the dipole emission rate in the bottom of Figure~\ref{fig:SIcomp_com_dip_quad}. Again, COMSOL (orange dots) and  Equation~\ref{Eq:rate_final} of the main text (solid line) are in agreement. Note that a fitting multiplication factor is used for the obtained rate since the rate depends on the distance between the two dipoles.

 \section{Square graphene nanoisland and selection rules}
 
\begin{figure*}[t]
     \centering
     \includegraphics[width=0.8\textwidth]{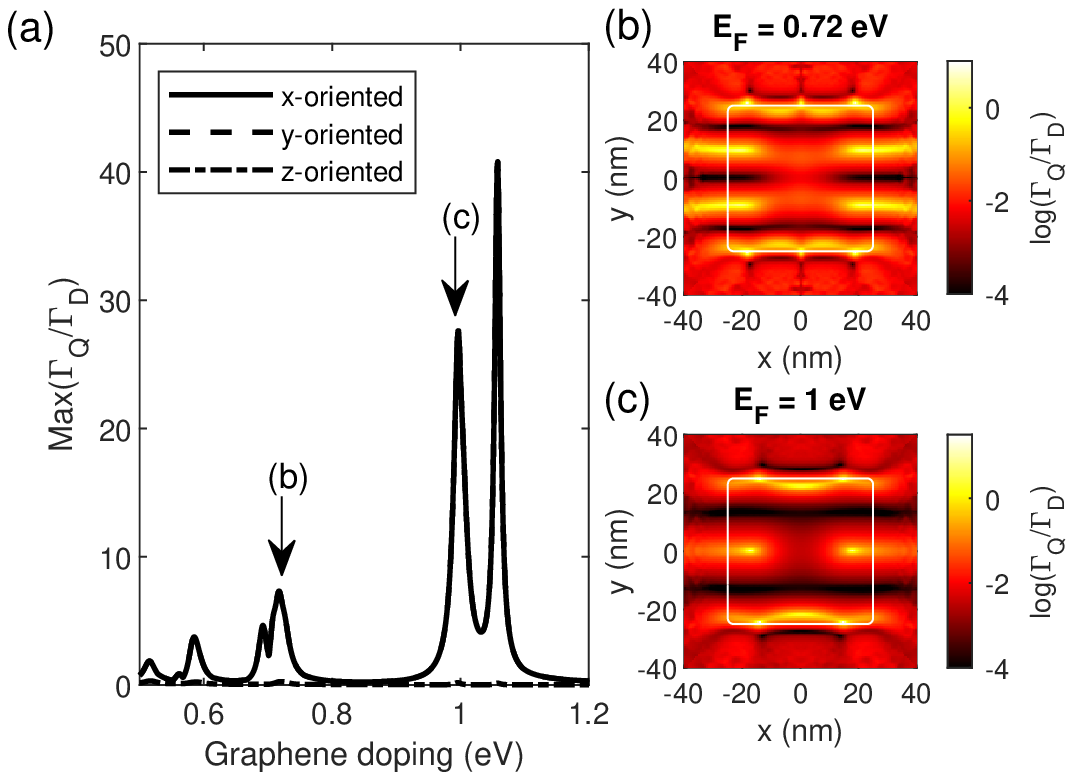}
     \caption{\textbf{Local breakdown of the selection rules with the square graphene nanoisland.} (a) The maximum of the quadrupolar rate ($\Gamma_Q$) over the dipolar rate ($\Gamma_D$) for a H-like emitter 5 nm above the square graphene nanoisland, evaluated for varying graphene doping and orientations of the emitter ($x$ and $y$-oriented emitters are superimposed). (b) The logarithmic value of the ratio for a doping $E_F=0.717$ eV and (c) $E_F=0.997$ eV. The boundaries of the square graphene nanoisland are represented with a solid white line.} \label{fig:SIQuadoverDip}
 \end{figure*}

 We focus on the square graphene nanoisland (50 nm side length), for which we computed the dipolar ($\Gamma_D$) and quadrupolar ($\Gamma_Q$) transition rates of an emitter 5 nm above its surface in Figure 1 in the main text. In Figure~\ref{fig:SIQuadoverDip}a, we plot the maximum of the ratio $\Gamma_Q/\Gamma_D$ over all positions of the emitter, as a function of the graphene doping, showing that the quadrupolar rate is up to 40 times stronger than the dipolar rate, breaking the conventional selection rules. The $x$ and $y$-oriented emitters are superimposed due to the symmetry of the square, while for the $z$-oriented emitter, the dipolar transition rate always dominates the quadrupolar transition rate.
 
Figures~\ref{fig:SIQuadoverDip}b and c show the ratio $\Gamma_Q/\Gamma_D$ in logarithmic scale for each position of the $y$-oriented emitter. The quadrupolar emission rate dominates along the edge of the nanoisland, where the field gradient is stronger. Note that it corresponds to a situation where the dipolar rate is relatively weakly enhanced compared to its maximum value ($\Gamma_D/\Gamma_{D,0} = 2\times 10^6$) : for example, at the maximum of the $E_F = 0.72$~eV situation, $\Gamma_D/\Gamma_{D,0} = 1.2\times 10^3$.

\section{Supplementary data: $y$- and $z$-oriented emitter}

\begin{figure*}[h]
     \centering
     \includegraphics[width=1.0\textwidth]{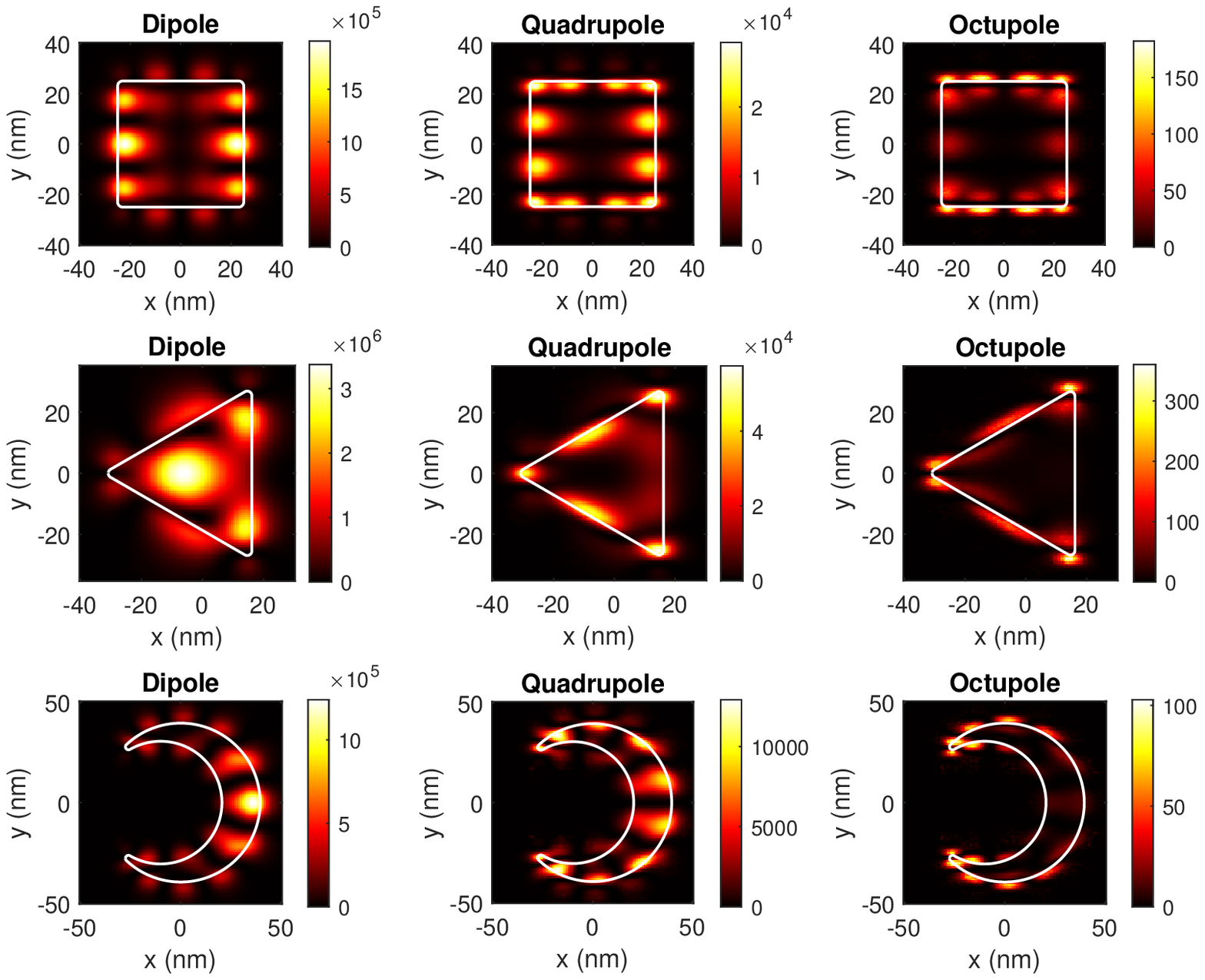}
     \caption{\textbf{Dipolar, quadrupolar and octupolar transition rates of a $y$-oriented emitter close to graphene nanoislands of various geometries.} The dipolar (left), quadrupolar (center), and octupolar (right) transition rates as a function of the emitter position 5 nm above the nanoisland: 50 nm side length square (up), 50 nm side length triangle (middle), and 80 nm height crescent (bottom). The geometry is represented with a solid white line, and the rates are normalized by the dipolar emission rate in free space $\Gamma_0$. For the triangular geometry, graphene is $E_F=0.98$~eV doped, for the square $E_F=0.72$~eV, and for the crescent $E_F=0.88$~eV.} \label{fig:SItrans_rates_y}
 \end{figure*}

\begin{figure*}[h]
     \centering
     \includegraphics[width=1.0\textwidth]{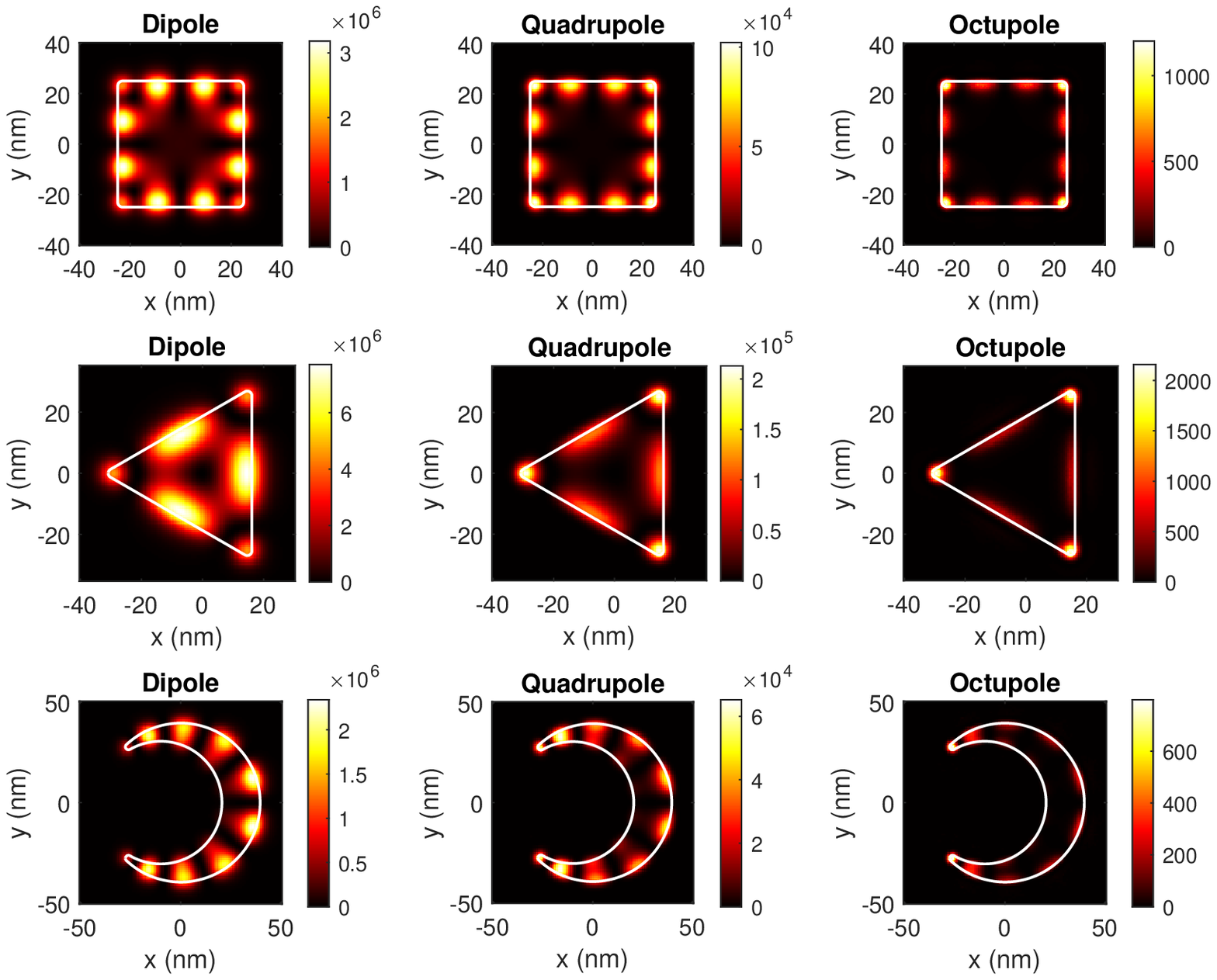}
     \caption{\textbf{Dipolar, quadrupolar and octupolar transition rates of a $z$-oriented emitter close to graphene nanoislands of various geometries.} The dipolar (left), quadrupolar (center), and octupolar (right) transition rates as a function of the emitter position 5 nm above the graphene nanoisland: 50 nm side length square (up), 50 nm side length triangle (middle), and 80 nm height crescent (bottom). The geometry is represented with a solid white line, and the rates are normalized by the dipolar emission rate in free space $\Gamma_0$. For the triangular geometry, graphene is $E_F=0.98$~eV doped, for the square $E_F=0.72$~eV,and for the crescent $E_F=0.88$~eV.} \label{fig:SItrans_rates_z}
 \end{figure*}

To complement Figure 1 of the main text, we plot here the dipolar, quadrupolar and octupolar transition rates of a $y$ (Figure~\ref{fig:SItrans_rates_y}) and a $z$ (Figure~\ref{fig:SItrans_rates_z}) oriented emitter 5 nm above square, triangle and crescent graphene nanoislands.

\end{document}